 \documentclass[prd,preprint,nofootinbib]{revtex4-2}
\usepackage{amssymb}
\usepackage{lipsum}
\usepackage{amsmath}
\usepackage{xcolor}
\usepackage{color}
\usepackage{graphicx}
\usepackage{soul}
\usepackage[T1]{fontenc} 
\setcitestyle{square} 
\usepackage[colorlinks=true, linkcolor=red, citecolor=blue, urlcolor=magenta]{hyperref}

\begin{document}
\title{On Deformed Phase Space Black Holes.}
\author{A. Crespo-Hern\'andez$^1$}
\email{andres.crespo@academicos.udg.mx}
\author{E. A. Mena-Barboza$^{2}$}
\email{eri.mena@academicos.udg.mx}
\author{M. Sabido$^{3,4}$}
\email{msabido@fisica.ugto.mx}
\affiliation{
$^{1}$Preparatoria 5, Universidad de Guadalajara Av. Fray Andr\'es de Urdaneta s/n, C.P. 44930, Guadalajara, Jalisco, M\'exico.\\
$^2$Centro Universitario de la Ci\'enega, Universidad de Guadalajara
Ave. Universidad 1115, Edif. Tutor\'ias e Investigaci\'on, C.P. 47820 Ocotl\'an, Jalisco, M\'exico.\\
$^{3}$Departamento  de F\'{\i}sica de la Universidad de Guanajuato, A.P. E-143, C.P. 37150, Le\'on, Guanajuato, M\'exico.\\
$^4$ Department of Physics, University of the Basque Country UPV/EHU, P.O BOX 644, 48080 Bilbao, Spain.
 }%
\date{\today}
\begin{abstract}
We revisit  the Schwarzschild black hole in the context of noncommutative phase space. By mapping the horizon area operator into a two-dimensional harmonic oscillator structure, we derive the modifications induced by phase space noncommutativity. The formalism is developed using the generalized Bopp shift on the canonical coordinates and momenta. We computed the modified area spectrum, black hole mass levels, transition frequencies, Hawking temperature, and entropy, highlighting the physical implications of the deformation parameters $\theta$ and $\eta$ on the thermodynamical properties of the deformed black hole. Finally, we conjecture a connection between the noncommutative contribution to the entropy and Barrow's entropy.
\end{abstract}
 \maketitle
\newpage
\section{Introduction}
Despite efforts to reconcile quantum field theory and General Relativity (GR), a satisfactory quantum theory of gravity has not been constructed. The search for a quantum theory of gravity, applying the well-known techniques of quantum field theory, yields a theory with ill-defined ultraviolet behavior, implying that the geometrical properties of spacetime change at the Planck scale. Moreover, because the scale at which the effects of quantum gravity are relevant for phenomenological exploration, there is no experimental guidance for the construction of the theory. Fortunately, 
if one assumes that black holes play a similar role in quantum gravity as the atom played in the development of quantum mechanics, we can use them to explore quantum aspects of gravity. Quantization of black holes was proposed in the pioneering work of Bekenstein\cite{bekenstein}. He suggested that the surface gravity is proportional to its temperature and that the area of its event horizon is proportional to its entropy and concluded that the horizon area should have a discrete spectrum with uniformly spaced eigenvalues. One physical consequence of the conjectured universal equal area spacing\cite{hod}, is that the quantum gravity description of the Schwarzschild black hole is characterized by a quantum area operator with equally spaced spectrum\cite{ahluwalia}.\\
The old idea of a noncommutative spacetime\cite{snyder} was revived in the context of string theory \cite{witten}. This idea was widely explored in particle physics and eventually studied in the context of gravity \cite{mia, Calmet, Wess,NC2}. As noncommutative effects are expected to be present near Planck's scale, one can consider an inherently noncommutative spacetime at the early ages of the universe. Unfortunately, using noncommutative gravity is a complicated ordeal due to the highly nonlinear character of these theories. Following the lessons from supersymmetric quantum cosmology, the effects of noncommutativity can be introduced by using the methods of noncommutative quantum mechanics \cite{gamboa} on the Wheeler-DeWitt (WDW) equation to construct noncommutative quantum cosmology\cite{Obregon2}. Moreover, the classical effects of noncommutative deformation were explored using the WKB approximation of the noncommutative quantum model \cite{eri1}. Although the noncommutative deformations of the minisuperspace were originally analyzed starting from the quantum level, ``classical'' noncommutative formulations were proposed \cite{Guzman:2008gz}. This construction allowed to introduce noncommutativity not only on the canonical variables but also on the canonical momentum. This deformed phase space cosmology was used to study the effects of the general deformation on different aspects of the Universe, in particular showing a late-time acceleration induced by the deformation.\cite{vakili, Miguel1,yee,huicho,shiraishi,rasouli,rasouli2,rasouli3,rasouli4}. {The initial studies of noncommutativity in black holes considered two fundamentally distinct approaches\footnote{After these early works on noncommutative black holes,  other approaches have been developed \cite{bastos,schneider,Touati:2022zbm,Banerjee:2009xx}.}. One considered that the effects of noncommutativity eliminated point-like sources for smeared objects \cite{nicolini}, replacing the mass of point particles with a Gaussian distribution and using the corresponding energy-momentum to find spherically symmetric and static solutions to GR. These solutions were interpreted as a noncommutative black hole, from which the thermodynamical properties were derived \cite{Banerjee:2008gc}. The other approach considered is the diffeomorphism between the Schwarzschild and Kantowski-Sachs (KS) metrics. The authors use the methods of noncommutative quantum cosmology on the WDW equation of the KS model to describe the noncommutative Schwarzschild black hole, allowing to study the effects of the noncommutative deformation on the thermodynamical properties of the noncommutative black hole \cite{ncbh}.}\\
The main goal of this work is to study the effects of the general noncommutative deformation on the Schwarzschild black hole and calculate the thermodynamical properties. Unlike the mentioned approaches, we will not use the diffeomorphism with the KS metric or replace point particles with a Gaussian distribution. {We will derive the spectrum of the deformed area operator, from which we obtain thermodynamic properties}. We find that temperature, entropy, and time of evaporation acquire corrections that appear to be of quantum origin and dependent on the noncommutative deformation parameters. Moreover, in the limit of the deformation on the momentum is removed, the result is consistent with noncommutative black holes  constructed using the WDW equation and the KS-Schwarzschild diffeomorphism. {Finally, we stablish a connection between the noncommutative correction of the entropy with Barrow's entropy.}

This paper is organized as follows. In Sec.~\ref{comutativo} we briefly review the quantization and calculate the spectrum for the commutative and deformed phase space black hole. In Sec.~\ref{termo}, we derive the thermodynamic properties. Lastly, Sec.~\ref{conclusions} is devoted to discussion and final remarks.

\section{Spectrum of the deformed black hole}\label{comutativo}

Let us begin by briefly reviewing the quantization of the thermodynamics of the Schwarzschild black hole\cite{Jalalzadeh:2022rxx}. Using the spherically symmetric Arnowitt–Deser–Misner (ADM) line element 
\begin{equation}
ds^2=-N^2dt^2+\Lambda^2(dr+N^rdt)^2+Rd\Omega^2,
\end{equation}
the canonical form of the Einstein-Hilbert action is
\begin{align}
&S=\int dt\int_{\Sigma_r}\left[\Pi_\Lambda\dot\Lambda+\Pi_R\dot R-NH-N^rH_r \right ]dr\\
&-\int\left[ N_+M_++N_-M_-\right]dt,\nonumber
\end{align} 
where $\Pi_\Lambda$ and $\Pi_R$ are the conjugate momenta of $(\Lambda,R)$, $\Sigma_r$ is the one dimensional Cauchy surface, with $H$ and $H_r$ are the super-Hamiltonian and radial super-momentum constraints. It has been shown that solving the constraints provides the observables\cite{Kuchar:1994zk}, allowing us to obtain the canonical conjugate momenta $P_M$ and rewrite the Hamiltonian in terms of the conjugate variables $(M,P_M)$. After choosing the time parameter, the right-hand-side asymptotic Minkowski time, the action reduces to
\begin{equation}
S=\int\left[P_M\dot M-H(M)\right]dt,
\end{equation}
where $H(M)=M$, is the reduced Hamiltonian. The Schwarzschild mass $M$ emerges as the only physical degree of freedom after imposing the Hamiltonian and momentum constraints. Its conjugate momentum $P_M$ is related to the coordinate of asymptotic time. Due to the Euclidean periodicity associated with the Hawking temperature $T_H$, which is subject to identification
\begin{equation}
	P_M \sim P_M + \frac{1}{T_H}.
\end{equation}
This periodic boundary condition implies that the phase space of $(M, P_M)$ is compact in the momentum direction. A convenient way to ``unwrap'' this structure is to perform a canonical transformation into variables $(\xi,p_\xi)$:
\begin{equation}\label{cantrans}
	\xi= \sqrt{\tfrac{A}{4\pi G}} \cos (2\pi P_M T_H), \quad
	p_\xi= \sqrt{\tfrac{A}{4\pi G}} \sin (2\pi P_M T_H),
\end{equation}
where $A = 16\pi M^2/M_P^4$ is the horizon area.
In these variables, the area takes the form
\begin{equation}
	A = 4\pi L_p^2 \left(\xi^2 + p_\xi^2\right), \label{Ac}
\end{equation} 
which is equivalent to the Hamiltonian of a harmonic oscillator. After canonical quantization of the WDW equation, one gets the equation for a harmonic oscillator and an equidistant area spectrum\cite{Jalalzadeh:2022rxx} given by
\begin{equation}\label{espectro}
  A_n = 8\pi L_P^2 \left(n + \tfrac{1}{2}\right),\quad
  M_n = \frac{M_P}{\sqrt{2}} \sqrt{n + \tfrac{1}{2}},
\end{equation} 
providing an equally spaced area levels and a discrete black hole mass spectrum. This area operator only captures the radial degrees of freedom, such as those associated with classical horizon and radial quantum fluctuations. To account for general quantum fluctuations and explore the microstructure of the black hole horizon, one can include ``internal'' phase space directions. A possible approach is to 
generalize Eq.(\ref{Ac}) to two dimensions\footnote{{Although the additional phase-space direction might  not correspond to independent physical degrees of freedom associated with the black hole horizon, the generalized phase space can be viewed as an effective description of the black hole system. 
From this perspective, these variables still carry physical significance, as they can provide a  parametrization of fluctuations of the effective  geometric observables.}}, 
with ${\xi}^2={x}^2+{y}^2$ and ${p}_\xi^2={p}_x^2+{p}_y^2$, leading to
\begin{equation}\label{canonico1}
		A = 4\pi L_p^2 \left( {p}_x^2 +{p}_y^2 + {x}^2 + {y}^2 \right).
\end{equation}
The solution for the polar coordinate is made considering
\begin{eqnarray}\label{polar_change}
	 	x &=& r \cos\phi, \quad p_x = p_r \cos\phi - \frac{p_\phi}{r} \sin\phi, \\
	 	y &=& r \sin\phi,\quad 
	 	p_y = p_r \sin\phi + \frac{p_\phi}{r} \cos\phi,\nonumber
\end{eqnarray}
and the operator A in polar coordinates becomes
\begin{equation}
{A} = 4\pi L_p^2 \left[ 
 p_r^2 + \frac{p_\phi^2}{r^2} + r^2 
\right],
\end{equation}
with the eigenfunction for the area operator given by
\begin{equation}\label{wavefunction}
	\psi_{n_r, l}(r, \phi) = \frac{1}{\sqrt{2\pi}} R_{n_r, l}(r) e^{i l \phi},
\end{equation}
where $l$ is the azimuthal quantum number. For the radial part $R_{n_r, l}(r)$ is obtained by substituting $\psi_{n_r, l}$ into the operator, then the effective radial equation is
\begin{equation}
	 -\frac{1}{2} \left[ \frac{d^2 R}{dr^2} + \frac{1}{r} \frac{d R}{dr} - \frac{l^2}{r^2} R \right] + \frac{1}{2} r^2 R = A' R,
\end{equation}
where $A'$ is an auxiliary eigenvalue related to the oscillator.  
The solutions for this are the usual Laguerre polynomials
\begin{equation}
	 	R_{n_r, l}(r)=\sqrt{\frac{2
    \,n_r!}{(n_r + |l|)!}}  \exp\left( -r^2 \right) L_{n_r}^{|l|}\left( r^2 \right)r^{|l|},
\end{equation}
where $L_{n_r}^{|l|}$ are the associated Laguerre polynomials, $n_r = 0, 1, 2, \dots$, $l = 0, \pm 1, \pm 2, \dots, \pm n$, and the normalization is $\int_0^\infty r |R_{n_r, l}(r)|^2 dr = 1$,
and after applying the $A$  operator on $\psi_{n_r, l}$, we get 
\begin{equation}\label{eq:A_polar}
	 A_{n_r,l} = 8\pi L_p^2 (2 n_r + |l| + 1). 
\end{equation}
representing the spectrum of the horizon area. 
This gives a degenerate spectrum for the radial direction, which is to be expected for a fluctuating horizon \cite{Bekenstein:1995ju,Kastrup:1997iu}. Moreover, for vanishing angular momentum $l=0$ one recovers Eq.(\ref{espectro}), this is expected because Eq.(\ref{Ac}) only describes the radial degrees of freedom.\\
In the deformed phase space approach, the deformation is introduced by the Moyal brackets $\{f,g\}_{\alpha}=f\star_{\alpha}g-g\star_{\alpha}f$, where the product between functions is replaced by the Moyal product
$(f\star{g})(x)=\exp{\left[\frac{1}{2}\alpha^{ab}\partial_{a}^{(1)}\partial_{b}^{(2)}\right]}f(x_{1})g(x_2)\vert_{x_1=x_2=x}$
such that
\begin{eqnarray}
\alpha =
\left( {\begin{array}{cc}
 \theta_{ij} & \delta_{ij}+\sigma_{ij}  \\
- \delta_{ij}-\sigma_{ij} & \eta_{ij}  \\
 \end{array} } \right),
\end{eqnarray}
where the $2\times 2$ matrices $\theta_{ij}$ and $\eta_{ij}$ are assumed to be antisymmetric and represent the noncommutativity in the coordinates and momenta, respectively. The resulting  $\alpha$ deformed algebra for the phase space variables is
\begin{align}\label{alg}
&\{x_i,x_j\}_{\alpha}=\theta_{ij}, \;\{x_i,p_j\}_{\alpha}=\delta_{ij}+\sigma_{ij},\; \{p_i,p_j\}_{\alpha}=\eta_{ij}.
\end{align}
An alternative  is to make a generalized Bopp shift  on the classical phase space variables $\{x,y,p_x,p_y\}$
\begin{eqnarray}\label{nctrans}
&\hat{x}=x-\frac{\theta}{2}p_{y}, \quad \hat{y}=y+\frac{\theta}{2}p_{x}, \\
&\hat{p}_{x}=p_{x}+\frac{\eta}{2}y, \quad \hat{p}_{y}=p_{y}-\frac{\eta}{2}x, \nonumber
\end{eqnarray}
where $\theta$ and $\eta$ are the noncommutative parameters. With these transformations, one recovers the modified algebra\footnote{In this work we have taken $\theta_{ij}=\theta\epsilon_{ij}$, $\eta_{ij}=\eta\epsilon_{ij}$ and $\sigma_{ij}=0$.}, this algebra is the same, but the Poisson brackets are different in the two algebras. For Eq.(\ref{alg}), the brackets are the $\alpha$ deformed ones and are related to the Moyal product; for the other algebra, the brackets are the usual Poisson brackets.

Now, we have to construct the modified phase space area operator. We start with the commutative area operator Eq.(\ref{canonico1}), but constructed with the variables that obey the modified algebra. After substituting these relations into the area operator, we get
\begin{align}
&\hat{A} = 4\pi L_P^2 \left[ \left(1+\tfrac{\theta^2}{4}\right)({p}_x^2+{p}_y^2) + \left(1+\tfrac{\eta^2}{4}\right)({x}^2+{y}^2)-(\theta+\eta)({x}{p}_y - {y} {p}_x) \right],
\end{align}
The first two terms correspond to an isotropic two–dimensional harmonic oscillator, while the last term is proportional to the angular momentum operator.
The operator $\hat{A}$ is the area operator in polar coordinates, which describes a two-dimensional harmonic oscillator with a coupling term proportional to the angular momentum $\hat{L}_z = x p_y - y p_x$.  
To find the spectrum for the new area operator, we use the change of variables in Eq.(\ref{polar_change}). In this variables $\hat{L}_z = -i  \frac{\partial}{\partial \phi}$.
and the operator $\hat{A}$ in  becomes	 
\begin{equation}
\hat{A} = 4\pi L_P^2 \left[ 
\Theta^2 \left( p_r^2 + \frac{p_\phi^2}{r^2} \right) 
+ \Xi^2 r^2 
- (\theta+\eta)\hat{L}_z 
\right],
\end{equation}
where 
$\Theta^2 = 1+\tfrac{\theta^2}{4}$ and   
$\Xi^2 = 1+\tfrac{\eta^2}{4}$,
are the new deformation quantities.
Then, after substituting Eq.(\ref{wavefunction}),
and defining $\gamma = \frac{\Xi}{\Theta}$, the effective radial equation becomes
\begin{equation}
	 -\frac{ \Theta^2}{2} \left[ \frac{d^2 R}{dr^2} + \frac{1}{r} \frac{d R}{dr} - \frac{l^2}{r^2} R \right] + \frac{1}{2} \Xi^2 r^2 R = A R,
\end{equation}
where $A$ is an auxiliary eigenvalue related to the ``oscillator".  
After the change of variable $u = \gamma r^2$, as before, the solutions are Laguerre polynomials and the eigenvalues of $\hat{A}$ are obtained by applying the operator to $\psi_{n_r, l}$.\\ The radial part of the oscillator has eigenvalues similar to those of the standard harmonic oscillator, with the frequency modified by $\Theta$ and $\Xi$. Also, because $\hat{L}_z \psi = l \psi$ for eigenstates with azimuthal quantum number $l$, the term $\hat{L}_z$ contributes $- (\theta + \eta) l$. Therefore, the spectrum has an extra contribution to the angular momentum arising from the deformation.  
Consequently, the effective radial operator corresponds to an oscillator with frequency $\omega = 2 \Theta \Xi$ with the spectrum 
\begin{equation}
	 A_{n_r, l} = 4\pi L_p^2 \left[
2 \Theta \Xi (2 n_r + |l| + 1) - (\theta + \eta) l
\right]. \label{eq:ANC_polar}
\end{equation}
These represent the quantized values of the horizon area, modified by the deformation parameters $\theta$ and $\eta$.
     
\section{Thermodynamics of the Deformed Black Hole}\label{termo}

In this section, we analyze the thermodynamic implications of the quantized noncommutative spectrum. The key idea is that radiation emission occurs through quantum transitions between neighboring energy levels. The frequency of the emitted quanta determines the Hawking temperature and, consequently, the entropy.

When describing black hole quantization, transitions between energy states play a central role. A black hole can move from a higher quantum level $n+1$, to a lower level $n$, releasing the difference as radiation \cite{mukhanov}. This process underlies  Hawking's radiation, in which the emitted quanta carry a discrete packet of energy. Importantly, such transitions suggest that complete evaporation does not occur; instead, a remnant of Planck-scale size remains once the process reaches its limit.

Hawking established that quantum fluctuations near the event horizon cause black holes to radiate thermally \cite{hawking}, with entropy proportional to one quarter of the horizon area. For a sufficiently massive black hole ($M \gg M_P$) and large quantum numbers ($n_r \gg 1$), the emission can be modeled as occurring when the system undergoes a spontaneous quantum jump. The characteristic frequency of radiation, often denoted $\omega_0$, directly reflects this quantized energy difference. Thus, the spectrum of emitted frequencies encodes information about the discrete nature of black hole states and the fundamental limits of evaporation.

To calculate the thermodynamic properties, we use the mass-spectrum operator. We will use the polar representation because it allows us to avoid some of the difficulties when working with the noncommutative deformation, since the noncommutative deformation contributes only to the angular part of the spectrum. 
Using Eq.\eqref{eq:A_polar} and the Bekenstein-Hawking area  $A_{BH}=16\pi M^2/M_p^2$ we find that the spectrum of the mass operator is
\begin{equation}
	M_{n_r,l} = \frac{M_p}{\sqrt{2}}\sqrt{2n_r+|l|+1
}.
\end{equation}
The frequency of emitted radiation $\omega_0$, when transitioning from the state $(n_r, l+1)$ to $(n_r, l)$ with $M>>M_p$ is given by

\begin{align}
&\omega_0(M)\equiv M_{n_r,l+1}-M_{n_r,l}\simeq \frac{M_p}{2\sqrt{2(2n_r+|l|)}}= \frac{M_p^{2} }{4M}\left[1+\left(\frac{M_{p}}{2M}\right)^2\right].
\end{align}

Now  we consider that the characteristic time scale before decay  is 
$\tau^{-1}_{n_r}=\vert\dot{M}\vert/\omega_0$, where $\vert\dot{M}\vert$ is the mass lost by evaporation\cite{Jalalzadeh:2022rxx,Xiang:2004sg}.  
{ 
The width of the state $W_{n_r,l}$} {is proportional to the distance between levels}\footnote{{In order to have a well defined structure with the condition $\epsilon\ll1$ as to be satisfied\cite{Mukhanov:1986me,mukhanov1990entropy}, as for $\epsilon>1$ we have overlapping levels. Moreover, the value of $\epsilon$ is fixed by requiring that the entropy corresponds to Bekenstein-Hawking entropy.}} { $W_{n_r,l}=\epsilon(M_{n_r,l+1}-M_{n_r,l})=\epsilon\omega_0$}. From the uncertainty relation $W_{n_r,l}\;\tau_{n_r}\approx1$  get
\begin{equation}
	\dot{M} = \epsilon \omega_0^2 \;\approx\; \epsilon \frac{M_p^4}{16M^2}\left(1 + \frac{ M_p^2}{2M^2}\right).
\end{equation}
Using the blackbody law $\dot{M}=\sigma_s A T^4$ where $A$ is the Bekenstein-Hawking area, $\sigma_s=\pi^2/60$ is the Stefan-Boltzmann constant, we obtain the following expression for the temperature
\begin{equation}
	T(M) \;\approx\; 
\left(\frac{ \epsilon}{\pi\sigma_s}\right )^{\frac{1}{4}}\frac{ M_p^2}{4 M}
\left(1 + \frac{ M_p^2}{8M^2}\right). \label{T_ab}
\end{equation}
The entropy follows from the relation $dS = dM / T$ integrated over $M$ and after taking $\epsilon=\sigma_s/(16\pi^3)$, we recover the entropy
\begin{equation}\label{S_ab}
	S=4\pi\frac{M^2}{ M_p ^2} - \frac{\pi}{2}\ln \left[4\pi\frac{M^2}{ M_p ^2}\right]+C
    = S_{HB}-\frac{\pi}{2}\ln S_{HB}+C.
\end{equation}
The leading term coincides with the Bekenstein–Hawking area law $S \propto M^2/M_p^2 $, whereas the logarithmic correction \cite{tkach} is a universal of virtually all approaches to quantum gravity arising from ultraviolet modifications. 

The evaporation time $\tau$ is obtained by integrating the mass loss equation $dt = dM / \dot{M}$ from initial mass $M_0$ to a remnant (approximated as $M \to 0$ for large $M_0$). So, the evaporation equation 
\begin{equation}
	\tau \;\approx\; 5120\pi\left(\frac{M_0^3}{M_p^4}\right) - 7680\pi\left(\frac{M_0}{M_p^2}\right), \label{tau_ab}
\end{equation}
Finally, the heat capacity $C = dM / dT$ to leading order is
\begin{equation}
	C \;\approx\; -\frac{8\pi M^2}{ M_p^2}\left(1 - \frac{3 M_p^2}{8M^2}\right).\label{C_ab}
\end{equation}
The negative leading term indicates instability; this is standard in black holes.\\
For the noncommutative case, we use the spectrum given in Eq.\eqref{eq:ANC_polar}, so the mass spectrum is given by
\begin{equation}
	M_{NC} = \frac{M_P}{2} \sqrt{2\Theta\Xi (2n_r + |l| + 1) - (\theta + \eta)l}\,.
\end{equation}	
The transition energy spacing becomes
\begin{equation}
  \omega_0 \approx \Theta\Xi\frac{M_P^2 }{4M}\left(1 - \frac{\theta + \eta}{2\Theta\Xi}\right)\left( 1 + \Theta\Xi\frac{M_P^2}{4M^2} \right), 
\end{equation}
{this approximation is valid\footnote{{This establishes  the scale for the values of the deformation parameters.}} for $\frac{\Theta\Xi M_p^2}{2M^2}\ll1$.  Following the same steps as in the commutative case} the mass loss rate is
\begin{equation}
	\dot{M}_{NC} =  \frac{\epsilon M_P^4}{16M^2}\left(1 - \frac{\theta + \eta}{2\Theta\Xi}\right)^2\left(1 +\Theta\Xi \frac{ M_p^2}{2M^2}\right).
\end{equation}
Following the same procedure as in the commutative case, the modified Hawking temperature reads
\begin{equation}\label{TNC_ab}
 T_{NC}= \frac{M_P^2}{4 M} \left(\frac{\epsilon}{\pi\sigma_s}\right)^\frac{1}{4}\sqrt{1 - \frac{\theta + \eta}{2\Theta\Xi}}\left( 1 + \Theta\Xi \frac{M_P^2}{2M^2} \right)^\frac{1}{4}.
\end{equation}
This reproduces the Hawking temperature $T_{\text{H}} \propto M_P/M$ in the limit $\Theta,\Xi \to 1$ (or equivalently $\theta,\eta\to0$), while noncommutative corrections reduce the temperature at the Planck scale. 
{In this way, we watch the contribution by noncommutativity, this is consistent with the standard Hawking temperature $T_H = M_p / (8\pi M)$}. 
Noncommutative corrections via $\Theta$ and $\Xi$, increase the radiation rate during the final stages of evaporation at small $M$, leading to {warmer} and shorter-lived remnant compared to the predictions of standard (commutative) general relativity.\\ 
Following the same approach as in the commutative case, we calculate the entropy 
\begin{align}\label{SNC_ab}
&S_{NC}  = \left(\frac{\pi\sigma_s}{\epsilon}\right)^\frac{1}{4}\frac{1}{\sqrt{1 - \frac{\theta + \eta}{2\Theta\Xi}}}\frac{2M^2}{M_P^2} -\left(\frac{\pi\sigma_s}{\epsilon}\right)^\frac{1}{4}\frac{\Theta\Xi}{4\sqrt{1 - \frac{\theta + \eta}{2\Theta\Xi}}}\ln\left(4\pi\frac{M^2}{M_p^2}\right)+const.
\end{align}
{Taking $\epsilon=\frac{\sigma_s}{16\pi^3\left(1 - \frac{\theta + \eta}{2\Theta\Xi}\right)^{2}}$ the entropy\footnote{{As in the commutative case the value of $\epsilon$ is fixed by requiring that the entropy corresponds to Bekenstein-Hawking entropy.}} can be written as} 
\begin{equation}\label{nc_entro}
   S_{NC}  = S_{BH}-\frac{\pi}{2}\ln \left(S^{\Theta\Xi}_{BH}\right), 
\end{equation}
{where} it is clear that the effects of the deformation are present at the quantum level. {Moreover,
in the commutative limit one recovers the standard Bekenstein-Hawking entropy $S_{BH} = 4\pi M^2 / M_p^2$ with the logarithmic correction.}\\ 
The other thermodynamic properties follow directly.  The noncommutative evaporation time is 
\begin{equation}
	\tau_{NC}\approx 5120\pi\frac{M_0^3}{M_p^4} - 7680\pi\;\Theta\Xi\left(\frac{M_0}{M_p^2}\right), \label{tauNC_ab}
\end{equation}
noncommutative terms shorten $\tau$ for large $M_0$, implying faster  evaporation due to enhanced phase space mixing { and consequently  disfavoring long-lived remnants. As this analysis remains purely thermodynamic and does not address the unitary evolution of the underlying quantum state, it does not provide a mechanism for information recovery  or a resolution to the BH information problem.}\\
Finally, the heat capacity $C = dM / dT$ to leading order is
\begin{equation}
	C_{NC}\approx -8\pi\frac{ M^2}{M_p^2}\left(1 - \Theta\Xi\frac{3 M_p^2}{8M^2}\right).\label{CNC_ab}
\end{equation}
As in the commutative case, a negative term appears, indicating instability as in standard black holes.

\section{Final Remarks}\label{conclusions}
In this paper, we have studied the thermodynamical properties of noncommutative black holes. The noncommutative deformation is introduced using the deformed phase space formulation, which is used in noncommutative cosmology.
After showing the redefinition of the phase space variables for the Schwarzschild black hole, we apply the Bopp shift to modify the Poisson algebra and construct a new deformed area operator and its corresponding  spectrum. From the spectrum, we define an effective angular frequency $\omega_0$ from which we calculate $\dot{M}$, and using the blackbody law $\dot{M}=\sigma_s A T^4$, we calculate the thermodynamic properties of the deformed black hole. As we have commented in section \ref{comutativo}, the calculations are performed using polar coordinates in order to interpret the noncommutative deformation as a contribution to the angular momentum. 
If we opted to calculate the spectrum using cartesian coordinates $(x,y)$. First, we define the creation and annihilation operators
\begin{align}
&a_{x_i}=\sqrt{\frac{\Xi}{2\hbar\Theta}}\,x_i+i\sqrt{\frac{\Theta}{2\hbar\Xi}}\,p_{x_i}, \quad a_{x_i}^\dagger=\sqrt{\frac{\Xi}{2\hbar\Theta}}\,x_i-i\sqrt{\frac{\Theta}{2\hbar\Xi}}\,p_{x_i},
\end{align}
where these operators satisfy $[a_{x_i},a_{x_j}^\dagger]=\delta_{ij}$. In this basis, the deformed area operator corresponds to a two-dimensional harmonic oscillator Hamiltonian with a noncommutative coupling term proportional to $\eta+\theta$, which mixes the $x$ and $y$ modes and prevents the area operator from being diagonal in the original Fock basis. However, one can perform a unitary transformation to a new set of ladder operators $(a'_x,a'_y)$ which diagonalize the quadratic form. 
In the new basis, the area operator is diagonal in the occupation numbers of $a'_x$ and $a'_y$. Consequently, the unitary transformation diagonalizes the noncommutative correction into two independent oscillators with shifted frequencies $4\pi L_p^2 [2\Theta\Xi \mp (\theta + \eta)]$.
Now we introduce the number operators $N_x' = a_x'^\dagger a_x'$, $N_y' = a_y'^\dagger a_y'$, which satisfy the eigenvalue equations   $N_{x_i}' |n_{x_i}', n_{x_j}'\rangle = n_{x_i}' |n_{x_i}', n_{x_j}'\rangle$ for $i\ne j$. 
Acting on the Fock basis $|n_x', n_y'\rangle$, and making $\nu_x=2\Theta\Xi - (\theta+\eta)$ and $\nu_y=2\Theta\Xi + (\theta + \eta)$,  
the deformed area operator is quantized in terms of the integers $n_x', n_y'$, but with a crucial asymmetry. The mode $x'$ has a frequency $\omega_{ox'}$ shifted by $4\pi L^2_p  \nu_x$, but the mode $y'$ has $\omega_{oy'}$ frequency $\omega_{ox'}$ shifted by $4\pi L_p^2  \nu_y$, thus  the deformation splits up the degeneracy of the two oscillators. Moreover,  for $\theta \to 0$ and $\eta \to 0$, the expression reduces to the symmetric form which corresponds to two identical harmonic oscillators with the same frequency $8\pi  L_p^2$. In this case, the degeneracy of the spectrum is restored, and we recover the equidistant quantization law that was previously derived\cite{Jalalzadeh:2022rxx}.
If we take  $\epsilon=\sigma_s/16\pi^3$ (as in the commutative case) when deriving the entropy, the first term of the entropy in Eq.(\ref{nc_entro}) is
\begin{equation}
S_{NC} = \frac{1}{\sqrt{1 - \frac{\theta + \eta}{2\Theta\Xi}}}S_{BH},
\end{equation}
a similar factor appears for the other thermodynamic properties. Therefore, to leading order, the noncommutative properties are proportional to the commutative ones. By taking $\eta=0$ and ${1 - \frac{\theta}{2\Theta}}=e^{2\sqrt{3}\tilde{\theta}p_{\gamma_0}}$, the first term in the entropy is the same as when using the KS WDW equation to describe the noncommutative black hole
.\\ After redefining Newton's constant, we can absorb the noncommutative parameter, but the noncommutative effects are not eliminated in the logarithmic contribution of the entropy. This is in contrast to the results when describing the noncommutative BH with KS WDW equation\cite{ncbh}. In that approach, if one redefines Newton's constant to such a way that the noncommutative parameter is not present, the noncommutative entropy has the same expression as the commutative entropy. Therefore, in this context, the effects of the noncommutative deformation can be eliminated by the redefinition of Planck's scale. In our case, after the redefinition that ``eliminates'' the deformation at the classical level, the noncommutative parameters are still present in the term involving the logarithmic correction. Because the logarithmic correction of the entropy is associated to quantum effects and considering that the corrections are derived from the quantum spectrum, one can conclude that the noncommutative contributions are relevant at the quantum level. Additionally,  after expanding $\Theta$ and $\Xi$ to leading order in $\theta$ and $\eta$, the noncommutative contribution can be separated from the commutative quantum contribution. \\
Let us now focus our attention on the logarithmic correction to the entropy. We start by  redefining the product $\Theta\Xi=1+\delta/2$  and find that $\delta\approx (\theta^2+\eta^2)/4$, and considering that $\theta$ and $\eta$ are small, we can take $0<\delta<1$. The argument of the logarithmic term (in terms of the area) has the form $ A^{1+\delta/2}$, which is Barrow's entropy\cite{Barrow:2020tzx}. This allows us to rewrite the entropy as
\begin{equation}
S_{NC}  = S_{BH}-\frac{\pi}{2}\ln S_{Barrow}.
\end{equation}
Writing the noncommutative entropy in this manner allows us to interpret the noncommutative deformation in the context of Barrow's entropy, and  because Barrow's entropy is related to the quantum~gravitational effects of the black hole horizon, we can infer that the noncommutative deformation is relevant at the quantum level. 
{Since Barrow's entropy is based on the assumption of a fractal, rather than a smooth horizon, the concept of a fuzzy horizon emerges. Moreover, noncommutativity introduces a  phase-space resolution that prevents the localization  below a minimal scale, leading to smeared horizon. Consequently, the entropy corrections induced by noncommutativity may effectively reproduce the area scaling associated with fractal horizons, suggesting that the Barrow exponent  can be interpreted phenomenologically as a measure of such microscopic noncommutative effects.}
Of course, a more precise understanding of the relationship between noncommutativity and Barrow's entropy is  desired, as it can provide a complementary or alternative description to the noncommutative deformation. This research is being conducted and will be reported elsewhere.
\section*{Acknowledgments}
{The authors would like to thank the anonymous referee who provided valuable comments which helped to improve the manuscript.} {\bf M. S.} is supported by the SECIHTI program  {\it ``Estancias sab\'aticas vinculadas a la consolidaci\'on de grupos de investigaci\'on''}. E. A. M. B. is partially supported by PEM (Pruebas El\'ectricas de M\'exico).
\bibliographystyle{unsrt}
\bibliography{ref}

@article{bekenstein,
    author = "Bekenstein, J. D.",
    title = "{The quantum mass spectrum of the Kerr black hole}",
    doi = "10.1007/BF02762768",
    journal = "Lett. Nuovo Cim.",
    volume = "11",
    pages = "467",
    year = "1974"
}

@article{Jalalzadeh:2022rxx,
    author = "Jalalzadeh, S.",
    title = "{Quantum black hole{\textendash}white hole entangled states}",
    eprint = "2203.09968",
    archivePrefix = "arXiv",
    primaryClass = "gr-qc",
    doi = "10.1016/j.physletb.2022.137058",
    journal = "Phys. Lett. B",
    volume = "829",
    pages = "137058",
    year = "2022"
}

@article{ahluwalia,
    author = "Ahluwalia, Dharam Vir",
    title = "{On quantum nature of black hole space-time: A Possible new source of intense radiation}",
    eprint = "astro-ph/9909192",
    archivePrefix = "arXiv",
    reportNumber = "ISGBG-01",
    doi = "10.1142/S0218271899000456",
    journal = "Int. J. Mod. Phys. D",
    volume = "8",
    pages = "651--657",
    year = "1999"
}

@article{hod,
    author = "Hod, Shahar",
    title = "{Bohr's correspondence principle and the area spectrum of quantum black holes}",
    eprint = "gr-qc/9812002",
    archivePrefix = "arXiv",
    doi = "10.1103/PhysRevLett.81.4293",
    journal = "Phys. Rev. Lett.",
    volume = "81",
    pages = "4293",
    year = "1998"
}

@article{mia,
    author = "Garcia-Compean, H. and Obregon, O. and Ramirez, C. and Sabido, M.",
    title = "{Noncommutative selfdual gravity}",
    eprint = "hep-th/0302180",
    archivePrefix = "arXiv",
    reportNumber = "CINVESTAV-FIS-03-08",
    doi = "10.1103/PhysRevD.68.044015",
    journal = "Phys. Rev. D",
    volume = "68",
    pages = "044015",
    year = "2003"
}

@article{Calmet,
    author = "Calmet, Xavier and Kobakhidze, Archil",
    title = "{Noncommutative general relativity}",
    eprint = "hep-th/0506157",
    archivePrefix = "arXiv",
    doi = "10.1103/PhysRevD.72.045010",
    journal = "Phys. Rev. D",
    volume = "72",
    pages = "045010",
    year = "2005"
}

@article{Wess,
    author = "Aschieri, Paolo and Dimitrijevic, Marija and Meyer, Frank and Wess, Julius",
    title = "{Noncommutative geometry and gravity}",
    eprint = "hep-th/0510059",
    archivePrefix = "arXiv",
    reportNumber = "DISTA-UPO-05, LMU-ASC-66-05, MPP-2005-199",
    doi = "10.1088/0264-9381/23/6/005",
    journal = "Class. Quant. Grav.",
    volume = "23",
    pages = "1883--1912",
    year = "2006"
}

@article{NC2,
    author = "Ohl, Thorsten and Schenkel, Alexander",
    title = "{Cosmological and Black Hole Spacetimes in Twisted Noncommutative Gravity}",
    eprint = "0906.2730",
    archivePrefix = "arXiv",
    primaryClass = "hep-th",
    doi = "10.1088/1126-6708/2009/10/052",
    journal = "JHEP",
    volume = "10",
    pages = "052",
    year = "2009"
}

@article{Obregon2,
    author = "Garcia-Compean, H. and Obregon, O. and Ramirez, C.",
    title = "{Noncommutative quantum cosmology}",
    eprint = "hep-th/0107250",
    archivePrefix = "arXiv",
    reportNumber = "CINVESTAV-FIS-55-01",
    doi = "10.1103/PhysRevLett.88.161301",
    journal = "Phys. Rev. Lett.",
    volume = "88",
    pages = "161301",
    year = "2002"
}

@article{gamboa,
    author = "Gamboa, J. and Loewe, M. and Rojas, J. C.",
    title = "{Noncommutative quantum mechanics}",
    eprint = "hep-th/0010220",
    archivePrefix = "arXiv",
    reportNumber = "USACH-FM-00-09",
    doi = "10.1103/PhysRevD.64.067901",
    journal = "Phys. Rev. D",
    volume = "64",
    pages = "067901",
    year = "2001"
}

@article{eri1,
    author = "Mena, E. and Obregon, O. and Sabido, M.",
    title = "{WKB-type approximation to noncommutative quantum cosmology}",
    eprint = "gr-qc/0701097",
    archivePrefix = "arXiv",
    doi = "10.1142/S0218271809014376",
    journal = "Int. J. Mod. Phys. D",
    volume = "18",
    pages = "95--106",
    year = "2009"
}

@article{Xiang:2004sg,
    author = "Xiang, Li",
    title = "{Black hole quantization, thermodynamics and cosmological constant}",
    doi = "10.1142/S0218271804004815",
    journal = "Int. J. Mod. Phys. D",
    volume = "13",
    pages = "885--898",
    year = "2004"
}

@article{vakili,
    author = "Malekolkalami, B. and Atazadeh, K. and Vakili, B.",
    title = "{Late time acceleration in a non-commutative model of modified cosmology}",
    eprint = "1411.3623",
    archivePrefix = "arXiv",
    primaryClass = "gr-qc",
    doi = "10.1016/j.physletb.2014.11.003",
    journal = "Phys. Lett. B",
    volume = "739",
    pages = "400--404",
    year = "2014"
}

@article{Miguel1,
    author = "P{\'e}rez-Pay{\'a}n, S. and Sabido, M. and Yee-Romero, C.",
    title = "{Effects of deformed phase space on scalar field cosmology}",
    eprint = "1111.6136",
    archivePrefix = "arXiv",
    primaryClass = "hep-th",
    doi = "10.1103/PhysRevD.88.027503",
    journal = "Phys. Rev. D",
    volume = "88",
    number = "2",
    pages = "027503",
    year = "2013"
}

@article{yee,
    author = "Sabido, M. and Yee-Romero, C.",
    title = "{Deformed phase space Kaluza{\textendash}Klein cosmology and late time acceleration}",
    eprint = "1511.08835",
    archivePrefix = "arXiv",
    primaryClass = "gr-qc",
    doi = "10.1016/j.physletb.2016.03.036",
    journal = "Phys. Lett. B",
    volume = "757",
    pages = "57--60",
    year = "2016"
}

@article{huicho,
    author = "L{\'o}pez, J. L. and Sabido, M. and Yee-Romero, C.",
    title = "{Phase space deformations in phantom cosmology}",
    eprint = "1711.01111",
    archivePrefix = "arXiv",
    primaryClass = "gr-qc",
    doi = "10.1016/j.dark.2017.12.006",
    journal = "Phys. Dark Univ.",
    volume = "19",
    pages = "104--108",
    year = "2018"
}

@article{shiraishi,
    author = "Kan, Nahomi and Aoyama, Takuma and Shiraishi, Kiyoshi",
    title = "{Classical and quantum bicosmology with noncommutativity}",
    eprint = "2206.10796",
    archivePrefix = "arXiv",
    primaryClass = "gr-qc",
    doi = "10.1088/1361-6382/aca868",
    journal = "Class. Quant. Grav.",
    volume = "40",
    number = "1",
    pages = "015010",
    year = "2023"
}

@article{rasouli,
    author = "Rasouli, S. M. M. and Ziaie, A. H. and Marto, J. and Moniz, P. V.",
    title = "{Gravitational Collapse of a Homogeneous Scalar Field in Deformed Phase Space}",
    eprint = "1309.6622",
    archivePrefix = "arXiv",
    primaryClass = "gr-qc",
    doi = "10.1103/PhysRevD.89.044028",
    journal = "Phys. Rev. D",
    volume = "89",
    number = "4",
    pages = "044028",
    year = "2014"
}

@article{rasouli2,
    author = "Rasouli, S. M. M. and Saba, Nasim and Farhoudi, Mehrdad and Marto, Jo{\~a}o and Moniz, P. V.",
    title = "{Inflationary Universe in Deformed Phase Space Scenario}",
    eprint = "1804.03633",
    archivePrefix = "arXiv",
    primaryClass = "gr-qc",
    doi = "10.1016/j.aop.2018.04.014",
    journal = "Annals Phys.",
    volume = "393",
    pages = "288--307",
    year = "2018"
}

@article{rasouli3,
    author = "Rasouli, S. M. M. and Farhoudi, Mehrdad and Khosravi, Nima",
    title = "{Horizon Problem Remediation via Deformed Phase Space}",
    eprint = "1006.5893",
    archivePrefix = "arXiv",
    primaryClass = "gr-qc",
    doi = "10.1007/s10714-011-1208-4",
    journal = "Gen. Rel. Grav.",
    volume = "43",
    pages = "2895--2910",
    year = "2011"
}

@article{rasouli4,
    author = "Rasouli, S. M. M. and Marto, Jo{\~a}o and Vargas Moniz, Paulo",
    title = "{Kinetic inflation in deformed phase space Brans{\textendash}Dicke cosmology}",
    eprint = "1805.05978",
    archivePrefix = "arXiv",
    primaryClass = "gr-qc",
    doi = "10.1016/j.dark.2019.100269",
    journal = "Phys. Dark Univ.",
    volume = "24",
    pages = "100269",
    year = "2019"
}

@article{ncbh,
    author = "Lopez-Dominguez, J. C. and Obregon, O. and Sabido, M. and Ramirez, C.",
    title = "{Towards Noncommutative Quantum Black Holes}",
    eprint = "hep-th/0607002",
    archivePrefix = "arXiv",
    doi = "10.1103/PhysRevD.74.084024",
    journal = "Phys. Rev. D",
    volume = "74",
    pages = "084024",
    year = "2006"
}

@article{Kuchar:1994zk,
    author = "Kuchar, Karel V.",
    title = "{Geometrodynamics of Schwarzschild black holes}",
    eprint = "gr-qc/9403003",
    archivePrefix = "arXiv",
    reportNumber = "UU-REL-94-3-1",
    doi = "10.1103/PhysRevD.50.3961",
    journal = "Phys. Rev. D",
    volume = "50",
    pages = "3961--3981",
    year = "1994"
}

@article{mukhanov,
    author = "Mukhanov, Viatcheslav F.",
    title = "{ARE BLACK HOLES QUANTIZED?}",
    journal = "JETP Lett.",
    volume = "44",
    pages = "63--66",
    year = "1986"
}

@article{hawking,
    author = "Hawking, S. W.",
    title = "{Black hole explosions}",
    doi = "10.1038/248030a0",
    journal = "Nature",
    volume = "248",
    pages = "30--31",
    year = "1974"
}

@article{tkach,
    author = "Obregon, O. and Sabido, M. and Tkach, V. I.",
    title = "{Entropy using path integrals for quantum black hole models}",
    eprint = "gr-qc/0003023",
    archivePrefix = "arXiv",
    doi = "10.1023/A:1010216126590",
    journal = "Gen. Rel. Grav.",
    volume = "33",
    pages = "913--919",
    year = "2001"
}

@article{Guzman:2008gz,
    author = "Guzman, W. and Sabido, M. and Socorro, J.",
    title = "{On Noncommutative Minisuperspace and the Friedmann equations}",
    eprint = "0812.4251",
    archivePrefix = "arXiv",
    primaryClass = "gr-qc",
    doi = "10.1016/j.physletb.2011.02.012",
    journal = "Phys. Lett. B",
    volume = "697",
    pages = "271--274",
    year = "2011"
}

@article{Banerjee:2008gc,
    author = "Banerjee, Rabin and Majhi, Bibhas Ranjan and Samanta, Saurav",
    title = "{Noncommutative Black Hole Thermodynamics}",
    eprint = "0801.3583",
    archivePrefix = "arXiv",
    primaryClass = "hep-th",
    doi = "10.1103/PhysRevD.77.124035",
    journal = "Phys. Rev. D",
    volume = "77",
    pages = "124035",
    year = "2008"
}

@article{nicolini,
    author = "Nicolini, Piero and Smailagic, Anais and Spallucci, Euro",
    title = "{Noncommutative geometry inspired Schwarzschild black hole}",
    eprint = "gr-qc/0510112",
    archivePrefix = "arXiv",
    doi = "10.1016/j.physletb.2005.11.004",
    journal = "Phys. Lett. B",
    volume = "632",
    pages = "547--551",
    year = "2006"
}

@article{witten,
    author = "Seiberg, Nathan and Witten, Edward",
    title = "{String theory and noncommutative geometry}",
    eprint = "hep-th/9908142",
    archivePrefix = "arXiv",
    reportNumber = "IASSNS-HEP-99-74",
    doi = "10.1088/1126-6708/1999/09/032",
    journal = "JHEP",
    volume = "09",
    pages = "032",
    year = "1999"
}

@article{snyder,
    author = "Snyder, Hartland S.",
    title = "{Quantized space-time}",
    doi = "10.1103/PhysRev.71.38",
    journal = "Phys. Rev.",
    volume = "71",
    pages = "38--41",
    year = "1947"
}

@article{bastos,
    author = "Bastos, Catarina and Bertolami, Orfeu and Costa Dias, Nuno and Nuno Prata, Joao",
    title = "{Non-Canonical Phase-Space Noncommutativity and the Kantowski-Sachs singularity for Black Holes}",
    eprint = "1012.5523",
    archivePrefix = "arXiv",
    primaryClass = "hep-th",
    doi = "10.1103/PhysRevD.84.024005",
    journal = "Phys. Rev. D",
    volume = "84",
    pages = "024005",
    year = "2011"
}

@article{schneider,
    author = "Schneider, Mathew and DeBenedictis, Andrew",
    title = "{Noncommutative black holes of various genera in the connection formalism}",
    eprint = "2005.01913",
    archivePrefix = "arXiv",
    primaryClass = "gr-qc",
    doi = "10.1103/PhysRevD.102.024030",
    journal = "Phys. Rev. D",
    volume = "102",
    number = "2",
    pages = "024030",
    year = "2020"
}

@article{Touati:2022zbm,
    author = "Touati, Abdellah and Zaim, Slimane",
    title = "{Thermodynamic properties of Schwarzschild black hole in non-commutative gauge theory of gravity}",
    eprint = "2204.01901",
    archivePrefix = "arXiv",
    primaryClass = "gr-qc",
    doi = "10.1016/j.aop.2023.169394",
    journal = "Annals Phys.",
    volume = "455",
    pages = "169394",
    year = "2023"
}

@article{Banerjee:2009xx,
    author = "Banerjee, Rabin and Gangopadhyay, Sunandan and Modak, Sujoy Kumar",
    title = "{Voros product, Noncommutative Schwarzschild Black Hole and Corrected Area Law}",
    eprint = "0911.2123",
    archivePrefix = "arXiv",
    primaryClass = "hep-th",
    doi = "10.1016/j.physletb.2010.02.034",
    journal = "Phys. Lett. B",
    volume = "686",
    pages = "181--187",
    year = "2010"
}

@article{Barrow:2020tzx,
    author = "Barrow, John D.",
    title = "{The Area of a Rough Black Hole}",
    eprint = "2004.09444",
    archivePrefix = "arXiv",
    primaryClass = "gr-qc",
    doi = "10.1016/j.physletb.2020.135643",
    journal = "Phys. Lett. B",
    volume = "808",
    pages = "135643",
    year = "2020"
}

@article{Bekenstein:1995ju,
    author = "Bekenstein, Jacob D. and Mukhanov, Viatcheslav F.",
    title = "{Spectroscopy of the quantum black hole}",
    eprint = "gr-qc/9505012",
    archivePrefix = "arXiv",
    doi = "10.1016/0370-2693(95)01148-J",
    journal = "Phys. Lett. B",
    volume = "360",
    pages = "7--12",
    year = "1995"
}

@article{Kastrup:1997iu,
    author = "Kastrup, H. A.",
    title = "{Canonical quantum statistics of an isolated Schwarzschild black hole with a spectrum E(n) = sigma n**(1/2) E(p)}",
    eprint = "gr-qc/9707009",
    archivePrefix = "arXiv",
    reportNumber = "PITHA-97-18",
    doi = "10.1016/S0370-2693(97)01121-0",
    journal = "Phys. Lett. B",
    volume = "413",
    pages = "267--273",
    year = "1997"
}

@article{Mukhanov:1986me,
    author = "Mukhanov, Viatcheslav F.",
    title = "{ARE BLACK HOLES QUANTIZED?}",
    journal = "JETP Lett.",
    volume = "44",
    pages = "63--66",
    year = "1986"
}

@article{mukhanov1990entropy,
  title={The entropy of black holes},
  author={Mukhanov, VF},
  journal={Complexity, Entropy, and the Physics of Information, Addison-Wesley},
  pages={47},
  year={1990}
}
\end{document}